# Design of GaAs-based valley phononic crystals with multiple complete phononic bandgaps at ultra-high frequency

Ingi Kim[1]*, Yasuhiko Arakawa[2], and Satoshi Iwamoto[1,2]

[1] *Institute of Industrial Science, University of Tokyo, Meguro, Tokyo 153-8505, Japan*
[2] *Institute for Nano Quantum Information Electronics, Meguro, Tokyo 153-8505, Japan*
*E-mail: kim-ingi@iis.u-tokyo.ac.jp

## Abstract

We report the design of GaAs-based monolithic valley phononic crystals (VPnCs) with multiple complete phononic bandgaps, which support simultaneous valley-protected edge states with different symmetries above GHz. Rotation of triangular holes in the unit cells breaks the mirror symmetry, and this orientation degree of freedom enables the structures to exhibit different valley vortex chiralities. We numerically demonstrate the transport of multi-band valley-protected edge states with suppressed backscattering at the sharp corners of the interfaces between different VPnCs. Such monolithic semiconductor structures pave the way for ultra-high frequency topological nanophononic applications by using the lithographic technique.

Topological band theory in condensed matter physics have received a lot of attention in the past decades[1,2]. Topological phenomena, such as quantum Hall (QH) effect, quantum spin Hall (QSH) effect, and topological insulators (TIs), have been observed in various electronic systems. One of the most remarkable properties of such topological phenomena is the emergence of topologically protected electron transport without backscattering from defects and imperfections. The presence of topological states is due to the non-trivial topological character of bulk electronic bands, which is called the bulk-edge correspondence[3,4]. Recently, topological concepts have been broadened to include classical systems such as photonic and phononic structures supporting the topologically protected states of light[5–14], acoustic[15–22], mechanical[23–26], and elastic[27–34] waves. According to the bulk-edge correspondence principle, the edge states of such classical waves emerge at the boundary between two structures that have topologically distinct phononic (or photonic) bandgaps. Most of the topological phononic structures in QH and QSH systems require complicated designs to achieve non-trivial topological phononic bandgaps. For example, additional active components, such as rotating gyroscopes or application of an external field, are needed to break the time-reversal symmetry in the QH systems[15,16,18,19,24,35]. The QSH systems require a double-Dirac cone achieved by the zone-folding method[31,33,34] or subwavelength perforations[28].

Another scheme to achieve topologically protected modes is to emulate the quantum valley Hall effect (QVHE). Valley is an additional degree of freedom (DOF) in momentum space of two-dimensional (2D) crystals with hexagonal point symmetries, such as graphene. Such crystals have a pair of degenerate states at the inequivalent K and K′ valleys in momentum space. The degeneracy at the K (K′) valley is lifted, which results in an energy gap with mirror symmetry breaking. The discrete valley DOF can be viewed as pseudospin, and the inter-valley scattering is suppressed because of the large separation in momentum. Based on the concept of the QVHE, the elastic valley-chiral edge states have been extensively studied in various phononic structures[36–44]. As the QVHE is realized by breaking only the lattice symmetries while preserving the time reversal symmetry, the complexity in design is significantly reduced compared with the elastic QH systems. It makes the topological structures less challenging to fabricate, especially when scaling down the design. At high frequencies, above MHz or GHz, valley phononic crystals (VPnCs) of continuous media are preferred in practical use. Recently, the VPnCs have been experimentally demonstrated in the continuous structures[37,39,45]. However, most of the studies on the VPnCs, including numerical studies, mainly focused on partial phononic bandgaps or certain elastic modes because of the design challenge in achieving complete phononic bandgaps due to the high modal density in the continuous structures. The VPnCs with compete phononic bandgaps are important to realize robust valley-protected transportation of elastic waves because defects can cause energy loss by coupling with other elastic modes in the partial bandgaps. Although the VPnC with a complete bandgap consisting of a hexagonal lattice of tungsten stubs deposited on a homogenous thin aluminum plate has been reported recently[43], monolithic design based on a single crystal may be preferred to realize the topological elastic waves at ultra-high frequency. Moreover, the VPnCs with multiple complete bandgaps of continuous medium are not fully explored despite potential practical applications of multi-functional topological devices.

In this paper, we present the design of a GaAs-based VPnC with multiple complete phononic bandgaps to realize valley-protected edge states for both $z$-symmetric (S) and anti-symmetric (A) modes at ultra-high frequency ranges. The complete bandgaps for each mode are simultaneously opened at different Dirac point frequencies near K (or K′) points by breaking the mirror symmetry. The Berry curvature characterizing the topological nature clearly shows opposite signs at the K and K′ valleys. The chiral characteristic of the bulk valley modes is controlled by rotating triangular holes in the unit cell through the topological phase transition. We numerically simulated the simultaneous valley-protected edge states for both S and A modes in their complete bandgaps. Finally, topologically protected valley transport of elastic waves was demonstrated to exhibit a suppressed backscattering at sharp corners. In contrast with the previously reported Si-based VPnCs with partial phononic band gaps[45], the piezoelectric effect of GaAs enables the excitation of ultrasonic elastic waves by applying an alternating voltage without additional piezoelectric transducers, as demonstrated in nanoelectromechanical systems[46,47]. With such monolithic semiconductor-based valley Hall elastic topological insulators, our design may contribute a significant step towards practical high-speed topological elastic devices, such as elastic isolation and radio frequency (RF) filters.

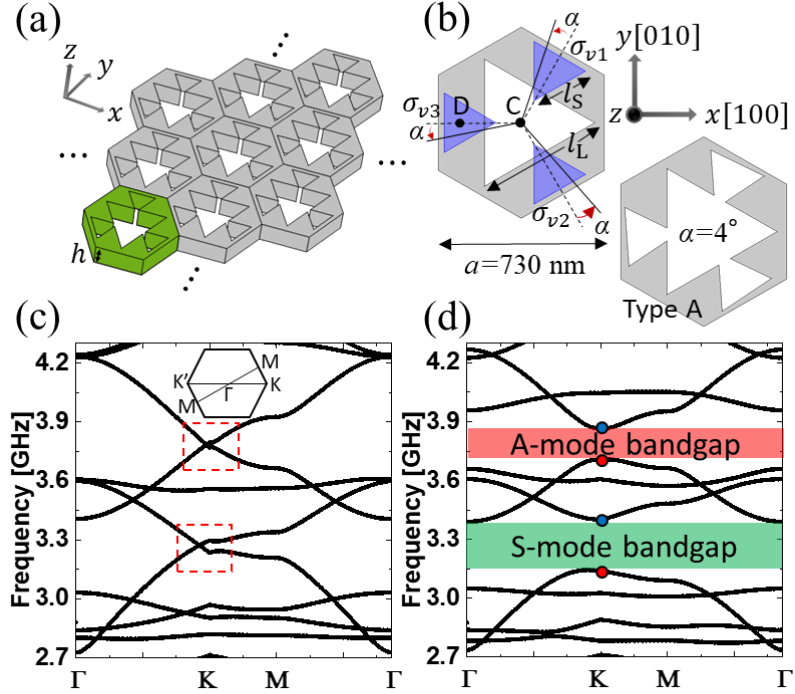

**Fig. 1.** (a) Schematic of the GaAs-based phononic crystal based on the triangular lattice; its primitive unit cell is shown in green. The thickness and period of the unit cell are $h$=200 nm and $a = 0.73$ μm, respectively. (b) Top view of the primitive unit cells with $C_{3v}$ symmetry (left) and broken mirror symmetry (right). The unit cells are formed by four regular triangular holes: large hole ($l_L$=535 nm) and three small holes ($l_S$=240 nm) surrounding the large hole. The positions of the small triangular holes are shown in blue; they are determined by the center-to-center distance $\overline{CD}$=260 nm, which allows them to slightly overlap with the large hole. The unit cell with broken mirror symmetry, which is referred as Type-A unit cell, is obtained by rotating only the small holes around the center of the unit cell by the angle $\alpha = 4°$. (c) and (d) Phononic band structures for the $C_{3v}$ symmetric unit cell ($\alpha = 0°$) and Type-A unit cell ($\alpha = 4°$), respectively. Two Dirac dispersions for S and A modes are formed near K (or K′) in different frequency ranges, as shown by red dashed lines in (c). The complete phononic bandgaps are simultaneously opened by breaking the mirror symmetry, as shown in (d). The different colors of the circles at K point indicate the vortex chirality of the valley states; namely, anticlockwise (red) and clockwise (blue) vortex patterns, as discussed in Fig. 2.

Figure 1 (a) shows a GaAs-based 2D phononic crystal patterned with two kinds of triangle-shaped holes arranged on a triangular lattice with the period $a = 730$ nm; its primitive unit cell is highlighted in green. The thickness $h$ of all studied phononic structures is 200 nm. The crystal directions [100] and [010] of GaAs correspond to the $x$ and $y$ axes, respectively, as shown in Fig. 1 (b). The sides of the large and small triangular holes are $l_L$=535 nm and $l_S$=240 nm, respectively. There are some overlaps between the large and small holes characterized by the center-to-center distance $\overline{CD}$=260 nm. By rotating small triangular holes (shown in blue) around the center of the unit cell, the mirror symmetry can be broken to reduce to $C_3$ symmetry from $C_{3v}$ symmetry. The vertical symmetric planes $\sigma_{v1}$, $\sigma_{v1}$, $\sigma_{v3}$ of the $C_{3v}$ symmetry are indicated by the dashed lines in Fig. 1 (b). Note that the large hole is fixed in its position to obtain two well-separated complete bandgaps through only the small perturbation. The unit cell with the rotation angle $\alpha = 4°$, which is referred as the Type-A unit cell, realizes one of the VPnCs with multiple complete bandgaps. Opposite rotation angle enables the structure to exhibit different topological properties. We additionally define the Type-B unit cell with $\alpha = -4°$ as the topologically different VPnC for the following discussions. The following material parameters are used for the numerical simulations using COMSOL Multiphysics, a commercial package based on the finite-element method: the mass density $\rho = 5360$ kg/m$^3$; the elastic constants $C_{11} = 118.8$, $C_{12} = 53.8$, and $C_{44} = 59.4$ GPa for GaAs[48]. The calculated band structures of the VPnCs with $\alpha = 0°$ and $\alpha = 4°$ (Type A) are shown in Figs. 1 (c) and (d), respectively. Two Dirac degeneracies at 3.26 and 3.77 GHz are highlighted with the red dashed-line squares near the high-symmetry point K in Fig. 1 (c). The perturbation that breaks $C_3$ symmetry but preserves $C_2$ symmetry simply shifts the Dirac cones without opening a gap[49]. Although the geometry of the unit cell possesses the $C_{3v}$ symmetry, it has such an anisotropic elastic property that the $C_3$ symmetry is effectively broken because of the elastic constant tensor of GaAs with cubic symmetry, while the $C_2$ symmetry is intact. This is why the Dirac points are slightly shifted from the K point, as shown in Fig. 1 (c). Because of the mirror plane perpendicular to the *z*-axis, all modes in the VPnCs can be distinguished into two sets by the symmetries of the *z*-displacement fields: *z*-symmetric (S) and anti-symmetric (A) modes. The first and second Dirac degeneracies are formed by the two bands for the S and A modes, respectively. The degeneracies are simultaneously lifted, which opens the complete phononic bandgaps by the broken mirror symmetry, as shown in Fig. 1 (d): the first bandgap from 3.14 to 3.39 GHz for the S-mode and the second bandgap from 3.7 to 3.87 GHz for the A-mode. The bandgap widths can be made larger by applying stronger mirror-symmetry breaking; for example, larger $\alpha$. However, large perturbation moves the other bands to the gap regions resulting in the partial bandgaps. Moreover, it can cause intervalley mixing, which ultimately removes the robustness of the edge modes against backscattering[39]. The rotation angle α and other geometric parameters that we chose allow the two complete bandgaps to exist simultaneously at different frequency ranges while suppressing intervalley mixing of the edge waves at the sharp corners, as will be discussed later.

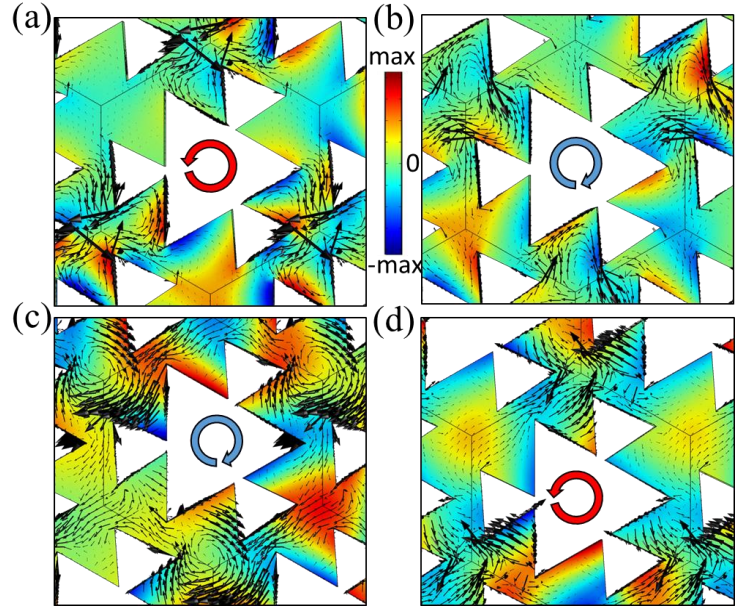

**Fig. 2.** Eigenstates of the Type-A VPnC at the valley K for S modes (a, b) and A modes (c, d) at lower and upper bands, respectively. The corresponding frequencies are marked with the red and blue circles in Fig. 1 (d). The red and blue curved arrows indicate different vortex chiralities corresponding to the anticlockwise and clockwise vortex patterns, respectively. The color scale and black arrows represent the amplitude of the z-displacement component and mechanical energy flux of the elastic waves, respectively.

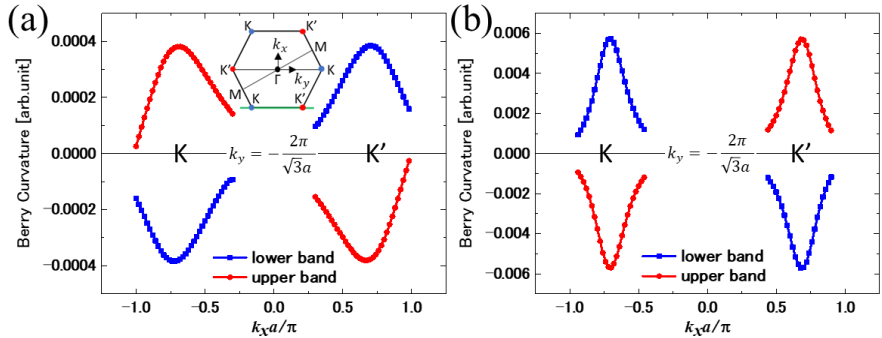

**Fig. 3.** Berry curvatures of the lower and upper bands for the S mode (a) and A mode (b) along the green line $k_y = -2\pi/\sqrt{3}a$ in the reciprocal space near the K and K′ valleys obtained from the numerical calculations in the case of the Type-A VPnC (α = 4°).

The broken mirror symmetry gives rise to the vortex chirality of the valley states for both the S and A modes at the high-symmetry K and K′ points. Figure 2 shows the z-displacement field patterns of the lower and upper valley states for the S and A modes at the K point in the case of the Type-A VPnC ($\alpha = 4°$). Figures 2 (a) and (b) represent the lower and upper states for the S modes; (c) and (d) represent the lower and upper states for the A modes, respectively. The frequencies of the valley states are marked by the circles in Fig. 1 (d). The mechanical energy flux shown in black arrows reveals the vortex chiralities, which are shown by the curved arrows in the centers of each figure. The fields at the K′ valley can be obtained by the time-reversal operation, which reverses the direction of the mechanical energy flux and, hence, the direction of the vortex pattern rotation[50]. Intervalley scattering can be suppressed by the quasi-angular momentum at the valleys and large separation in **k** space between the K and K′ points. To reveal the topological nature of the valley, Berry curvatures of both the S and A mode bands near the K and K′ points were numerically calculated along the $k_y = -2\pi/\sqrt{3}a$ line in the reciprocal space, as presented in Fig. 3. The signs of the Berry curvatures are opposite around K and K′ points because of the time reversal symmetry for both the S and A modes. Strong mirror symmetry breaking makes the distribution of Berry curvature less localized around the K and K′ points, which can result in weak reflections at sharp corner[39]. Noteworthily, the Berry curvatures for the S-mode bands have wider distribution than those for the A-mode bands, which are well localized near the K and K′ points, as shown in Figs. 3 (a) and (b), respectively. The reason is that the perturbation rotating the small triangular holes along *xy*-plane induces more deformation to the S-mode bands, in which the *x*- and *y*-components of the displacement field are dominant. The valley Chern numbers from such Berry curvatures are less than the expected quantized values $C_v = \pm\frac{1}{2}$ because of the overextension to the opposite symmetric point K (or K′) [39]. Although the QVHE in the designed VPnCs has weak topological properties, the edge states still show a substantially reduced backscattering, as will be discussed later.

The orientation DOF, which corresponds to the rotation of the small triangular holes, enables the structures to exhibit different valley vortex chirality through elastic valley Hall phase transition similarly to the other VPnCs[41,44]. To investigate valley-projected edge states by connecting two VPnCs with topologically distinct bandgaps, we considered another type of VPnC named Type B ($\alpha = -4°$), which possesses the opposite topological properties with respect to Type A. The topological characteristics of the Type-B VPnC were numerically calculated using the method discussed above. It is noteworthy that the Type-B structure can be viewed as the vertically-flipped Type-A structure. Indeed, one can easily notice that opposite chiral characteristics are observed by vertically inverting Fig. 2.

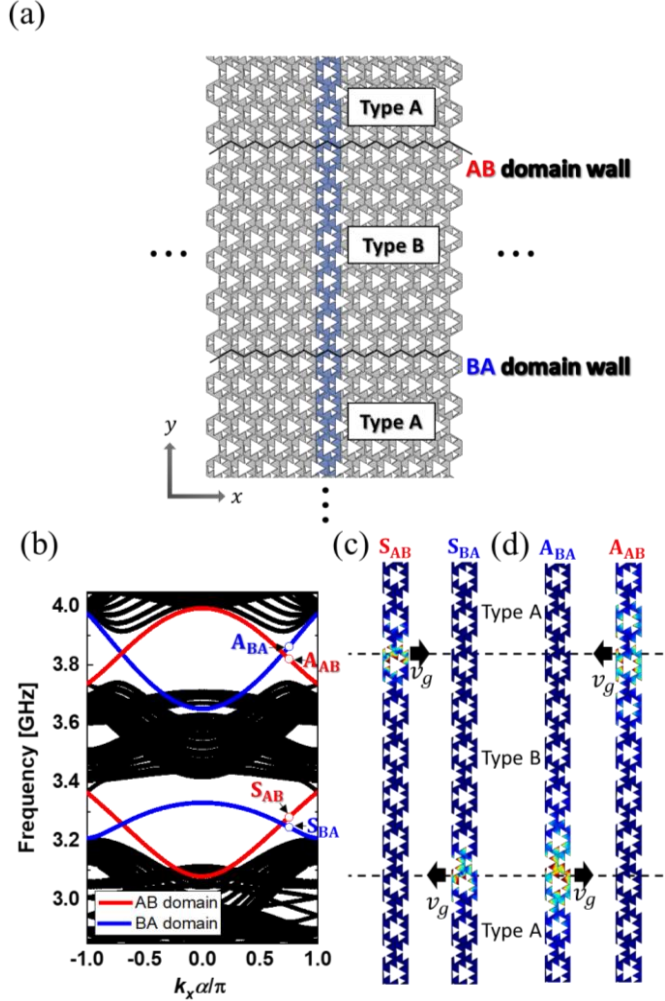

**Fig. 4** (a) Schematic of an elastic waveguide formed by connecting Type-A and Type-B VPnCs. This configuration creates two different domain walls, AB and BA, as marked by the solid black lines. The supercell of such a waveguide is shown in blue. (b) Calculated band structure of the supercell. The dispersion relations of the waveguide clearly show the existence of the valley-projected edge states for the S mode and A mode in the first and second complete bandgap regions, respectively. The red and blue lines indicate the edge states appearing at the AB and BA domain walls, respectively. (c, d) Plots of the eigenstates of the edge modes at the projected K valleys onto $k_x$ in the first and second bandgaps, respectively; the corresponding frequencies are marked with the circles in (b). The plots illustrate localized total displacement field patterns at the different domain walls with the directions of the group velocity.

To investigate the valley-projected edge states, Type-A and Type-B VPnCs were connected to form different domain walls, namely, AB and BA domain walls, as shown in Fig. 4 (a). According to the bulk-edge correspondence, the valley-projected edge states should appear along the interfaces. Figure 4 (b) shows the projected band structure as a function of $k_x$, the Bloch vector along the $x$ direction, which is parallel to the domain walls. The projected dispersion was calculated using the supercell indicated by the blue region in Fig. 4 (a). The edge states for the S and A modes are observed in the first and second complete bandgap regions, respectively. The red and blue lines indicate the two modes localized at the AB and BA domain walls, respectively. The two modes for each bandgap have opposite group velocities, but, as they are supported by the two different domain walls, they cannot couple. It is noteworthy that the edge states for the S and A modes appearing at each domain wall also have opposite group velocities. It originates from the different topological nature of the valley states for the S and A modes possessing opposite vortex chirality, as shown in Fig. 2. Total displacement fields of the edge states for the S and A modes clearly show their localization property, as shown in Figs. 4 (c) and (d), respectively.

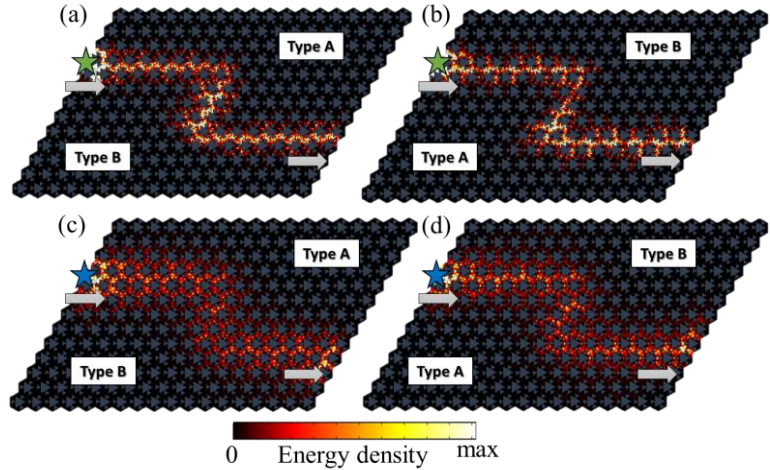

**Fig. 5** Full-field simulations of the valley-projected edge states for the S mode (a, b) and A mode (c, d) along the Z-shaped AB and BA domain walls. The color scale represents relative strain energy density of the elastic wave. The green and blue stars indicate the positions of the sources that are symmetric and antisymmetric along the $z$-axis, which excite the corresponding edge states, respectively. The propagation direction of the edge states are indicated by the gray arrows in each figure.

To investigate the robust propagation of the edge modes, full-field numerical simulations were performed. We calculated the steady-state responses of two different Z-shaped elastic waveguides along the AB and BA domain walls, as shown in Fig. 5. The strain energy density of each edge state is shown by the color scale. To reduce the impedance mismatch between the edge sates and excitation sources, the S-mode and A-mode edge states are excited by the symmetric and antisymmetric along the z-axis sources denoted by the green and blue stars, respectively. The A-mode edge states are less localized compared with the S-mode edge states. This feature originates from the smaller bandgap width of the A mode than that of the S mode, as shown in Fig. 1 (d), and not from the scattering or coupling with bulk modes. The less-localized features of the A-mode edge states are also consistent with the broaden mode shapes of the valley-projected edge states presented in Fig. 4 (d). Owing to the large separation between the K and K′ valleys in the momentum space and vortex chirality discussed above, both results clearly illustrate one-way propagation of the edge states along the domain walls with suppressed backscattering at the sharp corner.

To conclude, we propose the design of a GaAs-based VPnC with multiple complete bandgaps supporting valley-projected edge states for both the S and A modes. By breaking the mirror symmetry, two VPnCs with opposite topological properties were achieved. The bulk-edge correspondence reveals the existence of the edge states in the waveguides with different domain walls constructed by connecting two VPnCs. As the designed VPnC has the complete bandgaps, no other modes, which can cause loss by coupling with the edge states, exist in the bandgaps. Although the VPnCs have only weak topological properties, the valley projected edge states with substantially suppressed intervalley mixing were numerically demonstrated in the waveguides with sharp corners. Monolithic design of our VPnCs based on piezoelectric semiconductor materials is useful to control the topological edge states at ultra-high frequency above GHz and can be easily fabricated by using the lithographic technique. Thus, two channel edge states appearing at the complete bandgap regions have a great potential to realize high-performance and multi-functional phononic applications, such as signal processing in communication.

**Acknowledgments**

The authors would like to acknowledge Yasuhiro Hatsugai of University of Tsukuba, and Yasutomo Ota of University of Tokyo for fruitful discussions. This work was partially supported by MEXT KAKENHI Grant Number JP17J09077, JP15H05700, JP17H06138, JP15H05868